\documentclass{article}
\usepackage{spconf,amsmath,graphicx}
\usepackage{siunitx}
\usepackage{booktabs}
\usepackage{pgfplots,tikz}
\usetikzlibrary{positioning,fit,calc,shapes,matrix}
\usepackage{multirow}
\usepackage{cite}
\usepackage{url}
\usepackage{xcolor}
\usepackage{pifont}
\usepackage{etoolbox}

\def\L{{\cal L}}
\def\Lce{{\L^{(\text{CE})}}}
\def\Ltriplet{{\L^{(\text{TRIPLET})}}}
\def\Lresmask{{\L^{(\text{res-mask})}}}
\def\Lmse{{\L^{(\text{MMSE})}}}
\def\rsanblock{{NN}}

\def\DER{{DER}}
\def\SCONF{{SCER}}
\def\vect#1{\mathbf{#1}}
\def\mat#1{\mathbf{#1}}

\def\checkmark{\ding{51}}

\def\tblheading{\bfseries}
\def\emptycell{{---}}

\usepackage{hyperref}
\title{All-neural online source separation, counting, and diarization  \\for meeting analysis}
\name{
\begin{tabular}{c}	
Thilo von Neumann$^{1,2}$, Keisuke Kinoshita$^{1}$, Marc Delcroix$^{1}$, Shoko Araki$^{1}$, \\
Tomohiro Nakatani$^{1}$, Reinhold Haeb-Umbach$^{2}$
\end{tabular}
}
\address{
	$^1$ NTT Communication Science Laboratories, NTT Corporation, Kyoto, Japan\\
	$^2$ Paderborn University, Department of Communications Engineering, Paderborn, Germany\\
}

\definecolor{gray-box}{RGB}{105,105,105}
\tikzset{%
	every text node part/.style={align=center},
	>=stealth,
	block/.style={rectangle,draw=black!100,thick,minimum width=1.3cm,text height=1.5ex,text depth=.25ex,minimum height=0.4cm},
	branch/.style={circle,fill=black,minimum size=0.75mm,inner sep=0pt},
	arrow/.style={->,shorten >=0.01cm},
	operator/.style={circle,draw, minimum size=2.5mm,inner sep=0pt},
	node distance=0.5cm,
	box/.style={draw,gray-box,dashed,rounded corners},
	box-text/.style={fill=white,text=gray-box,inner sep=0}
}

\begin{document}
\ninept
\maketitle
\begin{abstract}
\vspace{-1mm}

Automatic meeting analysis comprises the tasks of speaker counting, speaker diarization,  and the separation of overlapped speech,  followed by automatic speech recognition. This all has to be carried out on arbitrarily long sessions and, ideally, in an online or block-online manner. While significant progress has been made on individual tasks, this paper presents for the first time an all-neural approach to simultaneous speaker counting, diarization and source separation. The NN-based estimator operates in a block-online fashion and tracks speakers even if they remain silent for a number of time blocks, thus learning a stable output order for the separated sources. The neural network is recurrent over time as well as over the number of sources. The simulation experiments show that state of the art separation performance is achieved, while at the same time delivering good diarization and source counting results. It even generalizes well to an unseen large number of blocks.


\end{abstract}
\vspace{-1mm}
\section{Introduction}
\vspace{-2mm}
\label{sec:intro}
The automatic  analysis of meetings promises to relieve humans from tedious transcription and annotation work. It comprises the tasks: (a) diarization, i.e.,  determining who is speaking when, (b) source counting, i.e., estimating the number of speakers in a meeting, (c) separating overlapped speech, i.e., carrying out (blind) source separation,  and (d) recognizing the separated streams. All of these are challenging tasks by themselves, which become even more demanding considering the fact that meetings can be arbitrarily long, making batch processing practically unfeasible and asking for block-online processing instead. 

In recent years, a substantial amount of research has been devoted to the meeting scenario\cite{Diarization_review, DIHARD_data, AMI_data}. 
One of the key challenges is the separation and recognition of overlapped speech. Perhaps surprisingly, even in professional meetings, the percentage of overlapped speech, i.e., time segments where more than one person is speaking, is in the order of 5\% - 10\%\footnote{measured on the AMI meeting corpus \cite{ AMI_data}.}, while in informal get-togethers it can  easily exceed 20\%\footnote{measured on the {\em Computational Hearning in Multisource Environments} (CHiME-5) database.}.
Recently, many promising neural network (NN)-based single-channel approaches have been proposed 
to solve the problem of source separation,
such as Deep Clustering (DC)\cite{Hershey_ICASSP16}, Deep Attractor Network (DAN)\cite{Chen_2016_arxiv}
and Permutation Invariant Training (PIT)\cite{Yu2016,Kolbaek2017}.
DC and DAN can be viewed as two-stage algorithms, where  in the first stage embedding vectors are estimated for each time-frequency (T-F) bin. In the second stage, these embedding vectors are clustered to obtain masks, from which the sources can be recovered by applying the masks to the speech mixture.  Note that the number of sources has to be known to determine the correct number of clusters. 
PIT, on the contrary, is a single-stage algorithm,  because it lets NNs directly estimate source separation 
masks without an explicit clustering step. In PIT, however, the network architecture depends on the maximum number of sources to be extracted. 

Considering this dependency on the number of sources, we proposed the Recurrent Selective Attention Network (RSAN), which is a purely NN-based mask estimator capable of, in theory, handling an arbitrary number of speakers \cite{RSAN}.
Specifically, RSAN is predicated  on a recurrent neural network (RNN) which can learn and determine how many iterations, i.e., source extraction processes, have to be performed to extract all sources \cite{Graves_2016_arxiv_adaptiveRNN}.
It extracts one source at a time from the mixture and repeats this process until all sources are extracted.
In experiments it achieved source separation performance comparable with PIT, and excellent source number counting accuracy.

None of these NN-based source separation algorithms \cite{RSAN, Hershey_ICASSP16,Chen_2016_arxiv,Yu2016,Kolbaek2017} 
has been extended to block or block-online processing in realistic situations, which consist of long recordings of an arbitrary number of intermittent speakers. Furthermore, a diarization component should be included, which ensures that the same speaker appears always at the same output node, even if he/she remains silent for some time.


Most conventional meeting diarization approaches perform 
block-offline or block-online processing 
by carrying out the following two steps sequentially \cite{Diarization_review, Araki_ICASSP2007, Araki_ICASSP2008, DIHARD_BUT, DIHARD_JHU}.
First, at each block, they perform separation (if necessary) and obtain speaker identity information
about each speaker in the block in the form of, e.g., i-vectors \cite{i-vector}, x-vectors \cite{x-vector}, 
or spatial signatures \cite{Araki_ICASSP2007, Araki_ICASSP2008, Drude_ICASSP2018}.
Then, the correct association of speaker identity information across  block boundaries, i.e., the  eventual diarization result, is established by
clustering this information in offline \cite{DIHARD_JHU,DIHARD_BUT} or online manners \cite{Araki_ICASSP2008}.
Here, block-offline processing is allowed to utilize future data, while the block-online processing is not.
In \cite{8169991}, joint separation and diarization is attempted using spatial mixture models. This, however, requires multichannel input and does not exploit spectral information for speaker re-identification.

Here, we also consider separation and diarization jointly, however proposing a novel all-neural block-online approach that performs source separation,
source number counting and diarization all together. 
The fact that the model is all-neural makes it possible to optimize the entire block-online process
through error back-propagation during NN training.
Importantly, in theory, the proposed method can handle any meeting situation
where, for example, a new speaker starts speaking in the middle of the meeting,
or one or more of the meeting attendees remain silent for a significant amount of time after his/her first utterances.
The method is an extension of \cite{RSAN}.

\section{Proposed Method}
\label{sec:prop}
\vspace{-2mm}


\subsection{Overall Structure}
\label{sec:prop:overall}
\vspace{-2mm}

\begin{figure}[t]
	\centering
	\input{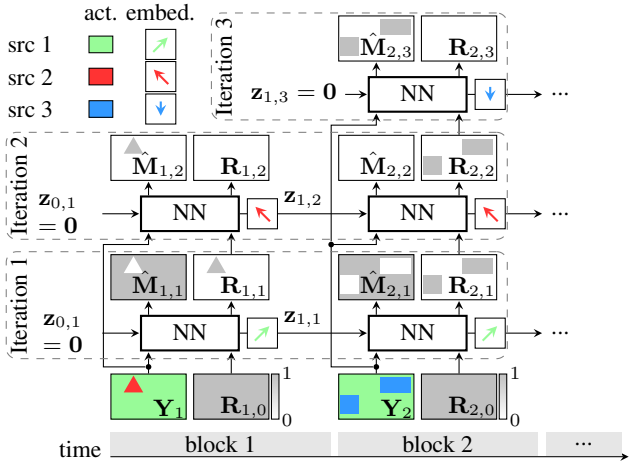}
	\caption[Proposed method unrolled over two time blocks and a maximum of three iterations]{
		Proposed method unrolled over two time blocks and a maximum of three iterations.
		In block 1, src1 \tikz{\node[legend-box,fill=color-src1]{};}, corresponding to backgound noise, and src2 \tikz{\node[legend-box,fill=color-src2]{};} are separated. Then, in block 2, the NN receives embedding vectors for src1 and src2, extracts src1, estimates an empty mask for silent src2, and extracts the new src3 \tikz{\node[legend-box,fill=color-src3]{};}.
	}
	\label{fig:overall}
\end{figure}

The algorithm works in a block-online manner and in each time block it successively extracts the sources until no sources are found anymore.
\autoref{fig:overall} depicts the processing steps for the first two time blocks and for up to three source extraction iterations per block.

Let $b$ denote the time block index and $i$  the iteration index within a block. 
At every iteration in a block, the network receives three inputs: the input spectrogram $\mat{Y}_b$, a residual mask $\mat{R}_{b,i-1}$ which is the output of the previous iteration on the same block, and a speaker adaptation input  $\vect{z}_{b-1,i}$ from the previous block. These inputs are processed in a neural network (``NN''), which outputs a source separation mask $\hat{\mat{M}}_{b,i}$, an updated residual mask $\mat{R}_{b,i}$, and a speaker embedding $\vect{z}_{b,i}$, which represents the identity of the extracted speaker.

The residual mask can be seen as an attention map that, once initialized with $\mat{R}_{b,0}=\mat{1}$ in every first iteration and updated in every following iteration, guides the network where to attend in order to extract a speaker that was not extracted in a previous iteration.
During test, the model decides when to stop the iterations based on a thresholding operation applied to the mean of the residual mask; 
it stops processing after iteration $i$, if the residual mask is virtually empty, i.e., $\frac{1}{TF} \sum_{tf} \left[\mat{R}_{b,i}\right]_{tf} < t_{\text{res-mask}}$.

In the first processing block, $b=1$, no speaker information is available from the previous block. Therefore, the input speaker information is set to zero: $\vect{z}_{0,i}=\vect{0}$. Without guidance, the network decides on its own in which order to extract the source signals.
The embedding vector $\vect{z}_{b,i}$ is passed as an adaptation input to the next time block, $b+1$, and guides the $i$-th iteration on that block to extract the same speaker as in $(b,i)$. This is related to the 'SpeakerBeam' concept to adapt a mask estimation network to a particular speaker\cite{SpeakerBeam}. Thus, it is ensured that all blocks extract the speakers in the same order.
In \autoref{fig:overall}, the different sources are indicated by their color, and it can be seen that the green source (src 1) is always extracted in the first,  the red in the second and the blue in the third iteration. If a source happens to be silent in a particular block (see the red source in block 2), then the mask is filled with zeros ($\hat{\mat{M}}_{b,i} = \vect{0}$), and the residual mask stays unmodified ($\mat{R}_{b,i} = \mat{R}_{b,i-1}$) in the iteration $i$ that is in charge of that source.

If the criterion to stop the speaker extraction iterations is not met after extracting all speakers found in previous blocks, 
the model increases the number of iterations to extract any new speaker (see iteration 3 of block 2 in \autoref{fig:overall}) until the stopping criterion is finally fulfilled.
To summarize, the network essentially attempts a guided source extraction for each source found in earlier blocks, 
and performs blind source separation on the remaining signal.

Note that the original RSAN\cite{RSAN} was formulated for processing only one block and does not receive and output speaker embeddings, 
whereas the proposed method, an extension of \cite{RSAN}, has enhanced capability of tracking speakers from block to block by doing so.

\vspace{-2mm}
\subsection{Details of used Neural Network}
\label{sec:prop:details-rsan-block}
\vspace{-2mm}

\begin{figure}[t]
	\centering
	\begin{tikzpicture}
	\node[block] (blstm) {BLSTM};
	\node[block,above=0.15cm of blstm] (linear1) {FC};
	\node[operator,above=4mm of linear1] (lhcu) {$\times$};
	\node[branch,above=0.5cm of lhcu] (branch2) {};
	
	\node[block,above=0.15cm of branch2] (linear-spk) {FC};
	\node[block,above=0.15cm of linear-spk] (time-avg) {time-avg};
	\node[block,above=0.15cm of time-avg] (affine) {affine};
	
	\node[block,left=0.3cm of linear-spk] (linear-mask) {FC};
	
	\node[block,left=0.3cm of lhcu] (linear-spk-in) {FC};
	\node[branch,left=0.2cm of linear-spk-in] (branch-in) {};
	
	\node[block,right=0.3cm of linear-spk] (linear-gate) {FC};
	\node[operator,right=0.3cm of linear-gate] (gate-mul) {$\times$};
	\node[operator,below=0.25cm of gate-mul] (gate-add) {$+$};
	\node[block] at (gate-add|-linear-spk-in) (norm) {norm};
	
	\node[box,yshift=1,fit=(linear-spk)(affine)] (box-spk) {};
	\node[box-text] at (box-spk.north) (box-spk-label) {\footnotesize spk. emb.};
	\node[box,fit=(linear-mask),xshift=-1] (box-mask) {};
	\node[box-text,rotate=90] at (box-mask.west) (box-mask-label) {\footnotesize mask};
	\node[box,yshift=1,fit=(linear-gate)(gate-mul)(gate-add)(norm)] (box-gate) {};
	\node[box-text] at (box-gate.north) {\footnotesize gate};
	\node[box,yshift=1,fit=(linear-spk-in)(lhcu)] (box-spk-adapt) {};
	\node[box-text] at (box-spk-adapt.north) {\footnotesize spk. adapt.};
	\node[box,fit=(blstm)(linear1),xshift=-1,minimum height=15mm] (box-blstm-fc) {};
	\node[box-text,rotate=90] at (box-blstm-fc.west) (box-blstm-fc-label) {\footnotesize BLSTM-FC};
	
	\node[block] at ($(affine.east)+(1.6cm,0.45cm)$) (rmc) {res-mask calculation};
	\node[branch] at ($(rmc -| linear-mask)+(0,2mm)$) (branch-mask-out) {};
	
	\draw[arrow] (branch-mask-out) -- (rmc.west|-branch-mask-out);
	\coordinate (k) at ($(norm.east)+(0.2cm,0)$) ;
	
	\node[above=0.3cm of rmc] (r-out) {$\mat{R}_{b,i}$};

	\coordinate[below = 0.08cm of blstm] (coord-concat);
	\coordinate  (concat-input-y) at ($(coord-concat)+(-4mm,-2mm)$);
	\coordinate  (concat-input-r) at ($(coord-concat)+(4mm,-2mm)$);


	
	\node at ($(concat-input-y-|linear-mask)-(0,5mm)$) (y) {$\mat{Y}_b$};
	\node at ($(concat-input-r-|rmc)-(0,5mm)$) (r) {$\mat{R}_{b,i-1}$};
	
	\draw (y) |- (concat-input-y) edge[in=270,out=0] (coord-concat);
	\draw (r) |-node[branch] (branch-r) {} (concat-input-r);
	\draw (concat-input-r) edge[in=279,out=180] (coord-concat);
	\draw[arrow] (coord-concat) -- (blstm);
	\draw[arrow] (branch-r) -| (k) -- (rmc.south-|k);

	\node[block,fit=(blstm)(box-mask)(box-spk)(box-gate)(box-spk-label)(box-mask-label)(rmc)(coord-concat)(branch-r)] {};
	
	\node[left=0.7cm of branch-in] (z-in) {$\vect{z}_{b-1,i}$};
	\node[right=0.5cm of norm] (z-out) {$\vect{z}_{b,i}$};
	\node at (linear-mask|-r-out) (mask-out) {$\vect{M}_{b,i}$};
	
	\draw[arrow] (blstm) -- (linear1);
	\draw[arrow] (linear1) -- (lhcu);
	\draw[arrow] (lhcu) -- (linear-spk);
	\draw[arrow] (branch2) -| (linear-mask);
	\draw[arrow] (linear-spk) -- (time-avg);
	\draw[arrow] (time-avg) -- (affine);
	\draw[arrow] (branch2) -| (linear-gate);
	\draw[arrow] (linear-gate) -- (gate-mul);
	\draw[arrow] (gate-mul) -- (gate-add);
	\draw[arrow] (gate-add) -- (norm);
	\draw[arrow] (branch-in) |- (gate-add);
	\draw[arrow] (branch-in) -- (linear-spk-in);
	\draw[arrow] (linear-spk-in) -- (lhcu);	
	\draw[arrow] (affine) -| (gate-mul);
	\draw (z-in) -- (branch-in);
	\draw[arrow] (norm) -- (z-out);
	\draw[arrow] (linear-mask) -- (mask-out);
	\draw[arrow] (rmc) -- (r-out);

	\end{tikzpicture}
	\vspace{-2mm}
	\caption{Detailed structure of the neural network. }
	\label{fig:detail}
\end{figure}
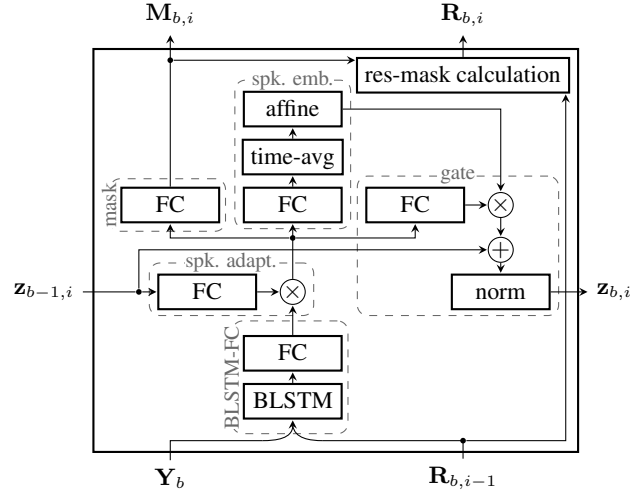

\autoref{fig:detail} depicts the detailed structure of the neural network ``NN'' in \autoref{fig:overall}.
In the figure, ``FC'' corresponds to a fully connected layer with a sigmoid activation,
``affine'' to affine transformation, ``time-avg'' to time averaging, and ``norm'' to length normalization.
The network consists of a common stack of bidirectional long short term memory (BLSTM) RNNs and a fully connected layer, hereafter denoted by BLSTM-FC, 
followed by multiple specialized parts (gray dashed boxes): 
A speaker adaptation network, a mask estimation network, a speaker embedding estimation network,  and a gate to control the update of speaker embedding vectors.

Two inputs to the NN, the spectrogram $\mat{Y}_b$ and the residual mask $\mat{R}_{b,i-1}$, are concatenated before being passed to the BLSTM-FC, and its output is passed through the speaker adaptation network, whose output is fed into the three remaining specialized networks.

Speaker adaptation is achieved by multiplying the transformed speaker adaptation input $\vect{z}_{b-1,i}$ with the activations from BLSTM-FC. 
By weighting the neurons based on the speaker embedding, the network behavior is modified to extract a specific speaker \cite{Delcroix_ICASSP2019}.

The speaker embedding estimation is inspired by 'Deep Speaker'\cite{Li2017}. Here, the output of a FC layer is averaged over time to condense the speaker information of the whole block $b$ into one embedding vector $\vect{z}_{b,i}$.
If a cosine distance-based loss function (described later) is used, this vector is further transformed and normalized. 

An optional gate is used to be able to pass the speaker embedding vector unmodified to the next time block, if  the speaker is silent in the current block. 
The gating mechanism ensures that the speaker information is unmodified if that speaker is absent in the current block, and updated if it is present. 


Finally, a source separation mask $\hat{\mat{M}}_{b,i}$ is estimated by the mask estimation network. This mask is used to update the residual mask by subtracting it from the residual mask obtained from the previous iteration and clipping to a range of $[0,1]$:
\vspace{-1mm}
\begin{equation}
\mat{R}_{b,i} = \max(\mat{R}_{b,i-1} - \hat{\mat{M}}_{b,i}, 0).
\end{equation}

\vspace{-3mm}
\subsection{Training Objectives}
\label{sec:prop:training}
\vspace{-2mm}
During training, the network is unrolled over multiple blocks and iterations and can be trained with back-propagation using the following multi-task cost function:
\begin{equation}
\mathcal{L} = \Lmse + \alpha \Lresmask + \beta \Lce + \gamma \Ltriplet,
\end{equation}
which is a weighted sum of the reconstruction loss $\Lmse$, the source counting loss $\Lresmask$, 
and the speaker embedding losses $\Lce$ and $\Ltriplet$.


The network is required to output a mask for a certain source at each iteration, but the order in which they will be extracted when they first appear, is not predictable.
Thus, a permutation-invariant loss function is required.
Once a source was extracted and the permutation was chosen to minimize the error on its first occurrence, its position is fixed for all following blocks. This is achieved, as explained earlier, by passing the embedding vectors from block to block.
Silent target masks $\mat{A}_{b,i}=\mat{0}$ are inserted when a source was active before, but is silent in the current block.

A permutation-invariant utterance-level mean square error (MSE) loss can be used as in RSAN\cite{RSAN}:
\begin{equation}
	\Lmse = \frac{1}{IB} \sum_{i,b} |\hat{\mat{M}}_{i,b}\odot \mat{Y}_b - \mat{A}_{\phi_b}|^2,
\end{equation}
where $I$ and $B$ are the total number of iterations and blocks, respectively. $\mat{A}$ is the target magnitude spectrogram.
The permutation $\phi_b$ for the $b$-th block is formed by concatenating the permutation used for the previous block $\phi_{b-1}$ with the permutation $\phi^*_b$ that minimizes the separation error for the
newly discovered sources in block $b$:
\begin{equation}
\phi_b = [\phi_{b-1}, \phi^*_{b}].
\end{equation}

To meet the  iteration stopping criterion, the following loss function is employed, 
that pushes the values of the residual mask to $0$ if no speaker is remaining\cite{RSAN}:
\begin{equation}
	\Lresmask = \sum_{b,tf} \left[\max\left(1-\sum_{i}\hat{\mat{M}}_{b,i}, 0\right)\right]_{tf}
\end{equation}

The speaker embedding vectors can be trained with a variety of loss functions.
Two possibilities are using an embedding layer followed by a softmax cross-entropy (CE) loss, hereafter called $\Lce$, and a triplet loss $\Ltriplet$\cite{Li2017}.

%

The triplet loss ensures  the cosine similarity between each pair of embedding vectors for the same speaker is greater than for any pair of vectors of differing speakers.
Triplets are formed by first choosing an anchor $\vect{a}$, and then for that anchor a positive $\vect{p}$ and a negative vector $\vect{n}$, which belong to the same and a different speaker than the anchor, respectively, from all embedding vectors of a minibatch.
Based on the cosine similarity $s_i^{\vect{an}}$ between the anchor and the negative,  and the cosine similarity $s_i^{\vect{ap}}$ between the anchor and the positive, the triplet loss for $N$ triplets can be formulated as
\begin{equation}
	\Ltriplet = \sum_{n=1}^{N} \max(s_n^{\vect{an}} - s_n^{\vect{ap}} + \delta, 0).
\end{equation}
where $\delta$ is a small positive constant.

\section{Experiments}
\label{sec:exp}
\vspace{-2mm}

\begin{table*}[t!]
	\centering
	\caption{SDR improvement, speaker diarization and speaker confusion error rates}
	\label{tbl:results}
		\def\unit#1{{[#1]}}
	\robustify\bfseries
	\sisetup{detect-weight=true,detect-inline-weight=math}
	\begin{tabular}{llllSSScSSScSSS}
			\toprule
		    &\multicolumn{3}{c}{\multirow{2}{*}{\tblheading Model}} & \multicolumn{3}{c}{(a) 4-block} && \multicolumn{3}{c}{(b) 12-block homogeneous} && \multicolumn{3}{c}{(c) 12-block conv.-like} \\
		     \cmidrule{5-7} \cmidrule{9-11} \cmidrule{13-15}
		    &&&& {SDR} & {DER} & {SCER} && {SDR} & {DER} & {SCER} && {SDR} & {DER} & {SCER} \\
		    && spk loss & gate & \unit{\si{\decibel}} & \unit{\si{\percent}} & \unit{\si{\percent}} && \unit{\si{\decibel}} & \unit{\si{\percent}} & \unit{\si{\percent}}  && \unit{\si{\decibel}} & \unit{\si{\percent}} & \unit{\si{\percent}} \\
		    \midrule
		    \multirow{4}{*}{\rotatebox[origin=c]{90}{Proposed}}&1 & \emptycell & \emptycell 			& \bfseries 19.4 & 4.2 & 3.1 && 7.5 & 5.5 & 5.3 && 11.5 & 7.8 & 6.5 \\
		    &2 & \emptycell & \checkmark 				& 18.5 & \bfseries 4.0 & \bfseries 2.6 && \bfseries 7.6 & \bfseries 5.2 & \bfseries 4.0 && 11.7 & \bfseries 6.6 & 4.9 \\
		    &3 & CE & \checkmark 	& 15.8 & 5.6 & 3.4 && 7.3 & 5.4 & 6.0 && 11.6 & 7.1 & 6.1 \\
		    &4 & triplet & \checkmark			& 17.9 & 4.2 & 2.9 && 7.2 & 5.6 & 4.8 && \bfseries 11.9 & 7.4 & 5.5 \\
		    \midrule
		    &\multicolumn{3}{l}{guess level}	& {\emptycell}   & 47.1 & 25.0 && \emptycell & 45.4 & 27.4 && \emptycell & 38.8 & 27.4 \\
		    &\multicolumn{3}{l}{ideal ratio mask (IRM)} & 28.9 & 0.8 & 0.0 && 14.2 & 1.0 & 0.0 && 24.0 & 0.8 & 0.1 \\
		    \midrule
		    \multirow{4}{*}{\rotatebox[origin=c]{90}{Baseline}}& i & \multicolumn{2}{l}{PIT batch} & 13.3 & 14.5 & 4.3 && 6.8 & 6.7 & 5.1 && 10.9 & 9.8 & \bfseries 4.4 \\
		    &ii & \multicolumn{2}{l}{RSAN batch} & 13.5 & 7.3 & 3.7 && 5.5 & 9.2 & 8.3 && 10.5 & 10.0 & 7.4 \\
		    &iii & \multicolumn{2}{l}{online clustering} & 8.2 & 15.8 & 9.8 && \emptycell & \emptycell & \emptycell && -10.0 & 52.1 & 37.8 \\
		    &iv & \multicolumn{2}{l}{offline clustering} & 10.9 & 9.7 & 4.5 && 3.2 & 17.5 & 7.9 && 5.3 & 15.8 & 6.2 \\
			\bottomrule
	\end{tabular}
\end{table*}

We evaluate the proposed method in terms of source separation and speaker diarization performance.
It is compared with two conventional methods and two simple extension of the conventional method for block processing:
(i) PIT and (ii) RSAN applied to the whole mixture (called PIT batch and RSAN batch, hereafter),
extensions of RSAN to perform diarization in (iii) block-online and (iv) block-offline manners.
These simple extensions are 2-stage methods similar to the conventional methods in \cite{Diarization_review, Araki_ICASSP2007, Araki_ICASSP2008, DIHARD_BUT, DIHARD_JHU},
which, based on NN, first separate the speakers, estimate associated speaker embedding vectors, 
and then cluster the vectors to estimate the correct association of speaker identity information among blocks.
The methods (iii) and (iv) are referred to as online and offline clustering, hereafter.
As the clustering method, we use a leader-follower and a hierarchical clustering algorithm for online and offline clustering, respectively.
While the offline clustering is performed with the correct number of speakers being given,
the other methods estimate it.
For reference purpose, we also show the performance of an oracle experiment using the ideal ratio mask (IRM), and a guess-level performance which assumes that exactly one speaker speaks all the time.
Throughout the experiments, the block size for all block processing schemes is set to 2.5 seconds, which amounts to about 150 time frames.

\subsection{Data}
\label{sec:exp:data}
\vspace{-2mm}

We generated meeting-like training and test data based on utterances taken from the single-channel WSJ0 corpus\cite{WSJ0}. 
To generate a speech mixture, one or two speech signals are mixed at a power ratio uniformly chosen between \SI{0}{\decibel} and \SI{5}{\decibel} relative to each other. Note, however, that the signals are not reverberant.

We created $55$ hours of training data, which were organized as a collection of $10$-second ($4$-block) mixtures.
Each mixture was generated such that the first \SI{5}{\second} contain a single or two speakers with a probability of \SI{50}{\percent} each,
while the second half contains silence/zero speakers, a single speaker or two speakers 
with a probability of \SI{15}{\percent}, \SI{55}{\percent} and \SI{30}{\percent}, respectively.

For evaluation, we generated $16$ hours of testing data in total,
which comprises (a) \SI{10}{\second} ($4$-block) mixtures whose utterance length and mixture characteristics match the training data, 
(b) \SI{30}{\second} ($12$-block) long homogeneous mixtures and (c) \SI{30}{\second} ($12$-block) long conversation-like mixtures.
The sets of speakers used for training and test are not overlapping.
The homogeneous and conversation-like mixtures are considerably longer than the training data, and thus can be used 
to test generalization capability of the proposed model.
In the homogeneous mixtures, one speaker talks throughout the whole mixture while another one starts speaking randomly 
in the first half of the mixture and continues speaking till the end.
Since there are no cases where a speaker stops speaking in the middle of the test data, 
the proposed model does not have to remember speakers over silent blocks.
The conversation-like mixtures are generated such that 
the first \SI{5}{s} of the test utterance contain a single or two speakers (\SI{50}{\percent} each),
while the mixture in the remaining time is generated such that it contains silence/zero, a single or two speakers 
with a probability of \SI{15}{\percent}, \SI{55}{\percent} and \SI{30}{\percent}, respectively.

\subsection{Network Configurations}
\label{sec:exp:netc-conf}
\vspace{-2mm}

Each neural network, including PIT and RSAN, had one fully connected layer on top of two BLSTM layers. 
This is the BLSTM-FC configuration referred to earlier.
The speaker embedding dimensionality was set to 128.
The speaker embedding estimation network consisted of $3$ fully connected layers with $50$, $50$ and $128$ neurons, respectively.
The weight for $\Lresmask$ was set to $\alpha=0.1$.

We employed 4 different architectures for the proposed method as in \autoref{tbl:results}: Model (1) does not use the gating mechanism while all other models (2), (3) and (4) do. Models (1) and (2) are trained only using the reconstruction and residual mask losses, without a speaker loss ($\beta=\gamma=0$).
Models (3) and (4) use the cross-entropy ($\beta=0.01, \gamma=0$) and triplet ($\gamma=0.1, \beta=0$) losses, respectively.

\subsection{Evaluation Metrics}
\label{sec:exp:eval-metric}
\vspace{-2mm}

We evaluate the performance in terms of signal-to-distortion ratio (SDR), 
diarization error rate (\DER) \cite{der}, and speaker confusion error rate (\SCONF).
DER indicates the percentage of time that the system outputs speech activity which is wrongly labeled:
\begin{equation}
	\DER = \frac{\text{\#frames with wrongly estimated speaker}}{\text{total \#frames}}\times 100\si{\percent}
\end{equation}
The error consists of missed speaker time (MST), false active time (FAT) and speaker error time (SET).
Note that if the system  confuses speakers, i.e., it correctly labels speakers as active, but confuses its output order,  then this is not considered as an error in \DER.
Therefore, SCER is additionally introduced as the percentage of time that the system confuses the output order of the speakers:
\begin{equation}
	\SCONF = \frac{\text{\#frames with confused speaker labels}}{\text{total \#frames}} \times 100\si{\percent}
\end{equation}
The number of frames with confused speaker labels is determined by comparing the optimal speaker assignment calculated for the whole mixture with the speaker assignment calculated for each frame.


\subsection{Results}
\label{sec:exp:results}
\vspace{-2mm}

As in \autoref{tbl:results}, the proposed methods outperform all four baselines i) - iv) in most cases in all three tested conditions in terms of SDR, DER and SCER.
The conventional two-stage methods, online clustering and offline clustering, failed to find correct association of speaker identity information among blocks,
and thus tend to work more poorly as the number of processed blocks increases.
As expected, PIT batch and RSAN batch worked significantly better than the two-stage methods.
However, interestingly, the proposed method generally worked better than these batch methods even though it performs block-online processing.

Model (2) of the proposed method outperforms model (1) in most scenarios, 
which shows the effectiveness of the gating function in \autoref{fig:detail}.
Again, interestingly, model (2) also outperforms model (3) and model (4) in almost all scenarios.
This suggests that the speaker embedding loss, be it $\Lce$ or $\Ltriplet$, actually disturbs the embedding process 
and the optimal adaptation vectors may contain additional information other than the speaker identity, e.g., about interfering signals.
Looking at a 2-dim. projection of the embedding space, we noticed that the embedding vectors form fairly condensed clusters for each speaker if an embedding loss is used, while it was not the case otherwise.
Last but not least, the models performed very well in source number counting (over \SI{98}{\percent} acc. in conv.-like data and over \SI{99}{\percent} acc. in all other cases), which is reflected in low \DER.
Some demos of the proposed method are available at \cite{demo_page}.

\section{Relation to prior works}
\label{sec:relation}
\vspace{-2mm}
Some researchers tried block-processing and online-processing based on NN-based source separation.
However,  
none of them has the capability of performing diarization in realistic situations 
where  speakers can stop talking in the middle of the meeting and remain silent for some time before they start talking again. 
In \cite{Yoshioka_Interspeech2018}, PIT is applied to each block, and then associations of estimated masks 
between adjacent blocks are estimated by a simple cross-correlation scheme. 
Clearly it cannot track speaker characteristics over silent blocks. 
A method proposed in \cite{Li_ICASSP2018} performs source separation in a frame-by-frame manner,  
by  exploiting  temporal  dependencies  and  continuity of  the  speech  signal. 
Specifically, a certain number of past frames of separated signals is used as additional input to a NN
which outputs separated signals for the current frame such that they can smoothly continue from past context data.
While it is similar to the proposed method in a sense that it utilizes speaker information appearing in the past,
it cannot deal with long silent regions since  it can see only a  limited past context of fixed length, e.g.,  600~ms in \cite{Li_ICASSP2018}.
On the other hand, the proposed method can naturally handle arbitrarily long silent regions, 
which we believe is a very important property when dealing with real meeting scenarios.

\vspace{-1mm}
\section{Conclusions}
\label{sec:conclusions}
\vspace{-2mm}

In this paper, we proposed an all-neural mask estimator which is capable 
of  block-online processing and which can adaptively change the number of output separation masks in each block.
It can track speakers even through silent blocks and detect new speakers in every block.
The experiments confirmed that the proposed method shows promising performance, both in terms of separation performance and in terms of diarization and speaker confusion error performance.

\vfill\pagebreak

\bibliographystyle{IEEEbib}
\bibliography{strings,refs}

\begin{thebibliography}{10}

\bibitem{Diarization_review}
X.~Anguera, S.~Bozonnet, N.~Evans, C.~Fredouille, G.~Friedland, and O.~Vinyals,
\newblock ``Speaker diarization: A review of recent research,''
\newblock {\em IEEE Transactions on Audio, Speech, and Language Processing},
  vol. 20, no. 2, pp. 356--370, Feb 2012.

\bibitem{DIHARD_data}
N.~Ryant, K.~Church, C.~Cieri, A.~Cristia, J.~Du, S.~Ganapathy, and
  M.~Liberman,
\newblock {\em First {DIHARD} Challenge Evaluation Plan}, 2018,
\newblock {https://zenodo.org/record/1199638}.

\bibitem{AMI_data}
J.~Carletta, S.~Ashby, S.~Bourban, M.~Flynn, M.~Guillemot, T.~Hain, J.~Kadlec,
  V.~Karaiskos, W.~Kraaij, M.~Kronenthal, G.~Lathoud, M.~Lincoln, A.~Lisowska,
  I.~McCowan, W.~Post, D.~Reidsma, , and P.~Wellner,
\newblock ``The {AMI} meeting corpus: A pre-announcement,''
\newblock in {\em The Second International Conference on Machine Learning for
  Multimodal Interaction, ser. MLMI'05}, 2006, pp. 28--39.

\bibitem{Hershey_ICASSP16}
J.~Hershey, Z.~Chen, J.~Le Roux, and S.~Watanabe,
\newblock ``Deep clustering: Discriminative embeddings for segmentation and
  separation,''
\newblock in {\em Proc. 2016 IEEE International Conference on Acoustics, Speech
  and Signal Processing (ICASSP)}, 2016, pp. 31--35.

\bibitem{Chen_2016_arxiv}
Z.~Chen, Y.~Luo, and N.~Mesgarani,
\newblock ``Deep attractor network for single-microphone speaker separation,''
\newblock in {\em Proc. 2017 IEEE International Conference on Acoustics, Speech
  and Signal Processing (ICASSP)}, March 2017, pp. 246--250.

\bibitem{Yu2016}
D.~Yu, M.~Kolb{\ae}k, Z.~Tan, and J.~Jensen,
\newblock ``Permutation invariant training of deep models for
  speaker-independent multi-talker speech separation,''
\newblock in {\em Proc. 2017 IEEE International Conference on Acoustics, Speech
  and Signal Processing (ICASSP)}, March 2017, pp. 241--245.

\bibitem{Kolbaek2017}
M.~Kolb{\ae}k, D.~Yu, Z.~Tan, and J.~Jensen,
\newblock ``Multitalker speech separation with utterance-level permutation
  invariant training of deep recurrent neural networks,''
\newblock {\em IEEE/ACM Transactions on Audio, Speech, and Language
  Processing}, vol. 25, no. 10, pp. 1901--1913, Oct 2017.

\bibitem{RSAN}
K.~Kinoshita, L.~Drude, M.~Delcroix, and T.~Nakatani,
\newblock ``Listening to each speaker one by one with recurrent selective
  hearing networks,''
\newblock in {\em Proc. 2018 IEEE International Conference on Acoustics, Speech
  and Signal Processing (ICASSP)}, April 2018, pp. 5064--5068.

\bibitem{Graves_2016_arxiv_adaptiveRNN}
A.~Graves,
\newblock ``Adaptive computation time for recurrent neural networks,'' 2016,
\newblock arXiv:1603.08983.

\bibitem{Araki_ICASSP2007}
S.~Araki, H.~Sawada, and S.~Makino,
\newblock ``Blind speech separation in a meeting situation with maximum {SNR}
  beamformers,''
\newblock in {\em Proc. 2007 IEEE International Conference on Acoustics, Speech
  and Signal Processing (ICASSP)}, April 2007, vol.~1, pp. I--41--I--44.

\bibitem{Araki_ICASSP2008}
S.~Araki, M.~Fujimoto, K.~Ishizuka, H.~Sawada, and S.~Makino,
\newblock ``Speaker indexing and speech enhancement in real meetings /
  conversations,''
\newblock in {\em Proc. 2007 IEEE International Conference on Acoustics, Speech
  and Signal Processing (ICASSP)}, April 2008, vol.~1, pp. 93--96.

\bibitem{DIHARD_BUT}
M.~Diez, F.~Landini, L.~Burget, J.~Rohdin, A.~Silnova, K.~Zmolikova,
  O.~Novotn{\'y}, K.~Vesel{\'y}, O.~Glembek, O.~Plchot, L.~Mo{\v s}ner, and
  P.~Mat{\v e}jka,
\newblock ``{BUT} system for {DIHARD} speech diarization challenge 2018,''
\newblock in {\em Proc. Interspeech 2018}, 2018, pp. 2798--2802.

\bibitem{DIHARD_JHU}
G.~Sell, D.~Snyder, A.~McCree, D.~Garcia-Romero, J.~Villalba, M.~Maciejewski,
  V.~Manohar, N.~Dehak, D.~Povey, S.~Watanabe, and S.~Khudanpur,
\newblock ``Diarization is hard: Some experiences and lessons learned for the
  {JHU} team in the inaugural {DIHARD} challenge,''
\newblock in {\em Proc. Interspeech 2018}, 2018, pp. 2808--2812.

\bibitem{i-vector}
N.~Dehak, P.~Kenny, R.~Dehak, P.~Dumouchel, , and P.~Ouellet,
\newblock ``Front-end factor analysis for speaker verification,''
\newblock {\em IEEE Trans. Audio, Speech, and Language Processing}, vol. 19(4),
  pp. 788--798, 2011.

\bibitem{x-vector}
D.~Snyder, P.~Ghahremani, D.~Povey, D.~Garcia-Romero, Y.~Carmiel, , and
  S.~Khudanpur,
\newblock ``Deep neural network-based speaker embeddings for end-to-end speaker
  verification,''
\newblock in {\em Proc. IEEE Spoken Language Technology Workshop}, 2016.

\bibitem{Drude_ICASSP2018}
L.~Drude, T.~Higuchi, K.~Kinoshita, T.~Nakatani, and R.~Haeb-Umbach,
\newblock ``Dual frequency- and block-permutation alignment for deep learning
  based block-online blind source separation,''
\newblock in {\em Proc. 2018 IEEE International Conference on Acoustics, Speech
  and Signal Processing (ICASSP)}, 2018, pp. 691--695.

\bibitem{8169991}
D.~Kounades-Bastian, L.~Girin, X.~Alameda-Pineda, R.~Horaud, and S.~Gannot,
\newblock ``Exploiting the intermittency of speech for joint separation and
  diarization,''
\newblock in {\em 2017 IEEE Workshop on Applications of Signal Processing to
  Audio and Acoustics (WASPAA)}, Oct 2017, pp. 41--45.

\bibitem{SpeakerBeam}
M.~Delcroix, K.~Zmolikova, K.~Kinoshita, A.~Ogawa, and T.~Nakatani,
\newblock ``Single channel target speaker extraction and recognition with
  speaker beam,''
\newblock in {\em Proc. 2018 IEEE International Conference on Acoustics, Speech
  and Signal Processing (ICASSP)}, 2018, pp. 5554--5558.

\bibitem{Delcroix_ICASSP2019}
M.~Delcroix, K.~Zmolikova, T.~Ochiai, K.~Kinoshita, S.~Araki, and T.~Nakatani,
\newblock ``Compact network for {SpeakerBeam} target speaker extraction,''
\newblock in {\em Proc. 2019 IEEE International Conference on Acoustics, Speech
  and Signal Processing (ICASSP)}, 2019,
\newblock (submitting).

\bibitem{Li2017}
C.~Li, X.~Ma, B.~Jiang, X.~Li, X.~Zhang, X.~Liu, Y.~Cao, A.~Kannan, and Z.~Zhu,
\newblock ``{Deep Speaker}: an end-to-end neural speaker embedding system,''
  2017,
\newblock arXiv:1705.02304v1.

\bibitem{WSJ0}
J.~Garofolo, D.~Graff, P.~Doug, and D.~Pallett,
\newblock {\em {CSR-I (WSJ0) Complete LDC93s6a}},
\newblock Linguistic Data Consortium, Philadelphia, New Jersey, 1993.

\bibitem{der}
NIST~Speech Group,
\newblock ``Spring 2007 (rt-07) rich transcription meeting recognition
  evaluation plan,'' 2007.

\bibitem{demo_page}
{{http://www.kecl.ntt.co.jp/icl/signal/kinoshita/publications/\\ICASSP19/online\_RSAN\_demo/index.html}}.

\bibitem{Yoshioka_Interspeech2018}
T.~Yoshioka, H.~Erdogan, Z.~Chen, X.~Xiao, and F.~Alleva,
\newblock ``Recognizing overlapped speech in meetings: A multichannel
  separation approach using neural networks,''
\newblock in {\em Proc. Interspeech 2018}, 2018, pp. 3038--3042.

\bibitem{Li_ICASSP2018}
Z.-X. Li, Y.~Song, L.-R. Dai, and I.~McLoughlin,
\newblock ``Source-aware context network for single-channel multi-speaker
  speech separation,''
\newblock in {\em Proc. 2018 IEEE International Conference on Acoustics, Speech
  and Signal Processing (ICASSP)}, May 2018, pp. 681--685.

\end{thebibliography}

\end{document}